\documentclass[a4paper,11pt]{article}
\usepackage[utf8x]{inputenc} 
\usepackage{jheppub}
\usepackage{amssymb}
\usepackage{braket}
\usepackage{graphicx}
\usepackage{hyperref}

\title{DAMPE Electron-Positron Excess in Leptophilic $Z'$ model}

\author[a]{Karim Ghorbani }
\author[b]{and Parsa Hossein Ghorbani}

\affiliation[a]{Physics Department, Faculty of Sciences, Arak University, Arak 38156-8-8349, Iran}
\affiliation[b]{Institute for Research in Fundamental Sciences (IPM), 
 School of Particles and Accelerators, P.O. Box 19395-5531, Tehran, Iran}

\emailAdd{karim1.ghorbani@gmail.com}
\emailAdd{parsaghorbani@gmail.com}


\abstract
{Recently the DArk Matter Particle Explorer (DAMPE) has reported an excess in the electron-positron flux 
of the cosmic rays which is interpreted as a dark matter particle with the mass about $1.5$ TeV. We come
up with a leptophilic $Z'$ scenario including a Dirac fermion dark matter candidate 
which beside explaining the observed DAMPE excess, is able to 
pass various experimental/observational constraints including the relic density value from the WMAP/Planck, 
the invisible Higgs decay bound at the LHC, the LEP bounds in electron-positron scattering, 
the muon anomalous magnetic moment constraint, Fermi-LAT data,
and finally the direct detection experiment limits from the XENON1t/LUX. By computing the 
electron-positron flux produced from a dark matter with the mass about $1.5$ TeV 
we show that the model predicts the peak observed by the DAMPE. 
}

\begin{document}

\maketitle

\section{Introduction}
One of the signals of a new physics could be the observation of any excess in the energy spectra of the cosmic rays.
The search for such an excess in the electron and positron spectra have been already in progress by different particle 
detectors in the space; the PAMELA satellite experiment observed an abundance of the positron in the cosmic radiation 
energy range of
$15-100$ GeV \cite{Adriani:2008zr}, also a positron fraction in primary cosmic rays of $0.5-350$ GeV \cite{Aguilar:2013qda}
and $0.5-500$ GeV \cite{Accardo:2014lma} and the measurement of electron plus positron flux in the primary 
cosmic rays from $0.5$ GeV to $1$ TeV \cite{Aguilar:2014fea} reported by the Alpha Magnetic Spectrometer (AMS02). 
The motivation of the current paper is however the recent report of the first results of 
the DArk Matter Particle Explorer (DAMPE) with unprecedentedly high energy resolution and low background in the 
measurement of the cosmic ray electrons and positrons (CREs) in $25$ GeV to $4.6$ TeV energy range \cite{Ambrosi:2017wek}.  
At energy about $1.4$ TeV a peak associated to a monoenergetic electron source is observed. This excess is interpreted by
a dark matter particle with the mass around $1.5$ TeV annihilating into electron and positron in a nearby 
subhalo in the Milky Galaxy about $0.1-0.3$ kpc distant from the solar system. 
The dark matter annihilation cross section times velocity is estimated to be in the range $\sim 10^{-26}-10^{-24}$ cm$^3$/$s$ for 
the aforementioned dark matter mass. For an interpretation of the DAMPE data see \cite{Yuan:2017ysv}.

There are already several papers that have tried to explain this excess using different models.
In \cite{Fan:2017sor} a vector-like fermion DM with a new U(1) gauge boson which only couples to the first two
lepton generation is used to explain the DAMPE data.
In this direction, model independent analysis performed with
fermion DM in \cite{Duan:2017pkq} and with scalar and fermionic DM in \cite{Athron:2017drj}.
There are also studies within the simplified models with a $Z^\prime$ gauge bosons couples only to the first
family of leptons (electrophilic interaction) or to the other families as well \cite{Gu:2017gle,Chao:2017yjg,Cao:2017ydw,Liu:2017rgs}.
There is another study in \cite{Chao:2017emq} where electron flavored fermion DM can interact with the first
generation lepton doublet via an inert scalar doublet or with right-handed electron via a charged scalar singlet.
In addition, the excess is studied in Hidden Valley model with lepton portal DM \cite{Tang:2017lfb},
radiative Dirac seesaw model \cite{Gu:2017bdw} and gauged $L_e - L_\mu$ model  \cite{Duan:2017qwj}.
It is also studied that the  DM particles annihilate to two intermediate scalar particles and then
the scalars decay to DM fermions \cite{Zu:2017dzm}.
In \cite{Gao:2017pym} it is shown that a DM candidate with cascade decay can explain the DAMPE
TeV electron-positron spectrum. 
There are detailed analysis on the morphology of CRE flux considering properties of the primary
electron sources \cite{Huang:2017egk,Jin:2017qcv,Yang:2017cjm}. 

Meanwhile, it should be noted that there may exist some possible exotic sources for the excess
or it may originate from some standard sources like pulsars or supernova remnants.
In this work we interpret the excess due to the DM annihilation in a nearby halo. 

To explain the DAMPE excess, we come up with a leptophilic $Z'$ dark matter scenario that contains a Dirac fermion which 
plays the role of the dark matter candidate. Besides, in the dark sector we introduce a $U(1)'$ gauge symmetry and a 
complex scalar that together with the Dirac fermion are charged under this $U(1)'$ gauge symmetry. The dark sector 
communicates with the standard model sector through two portals. One portal is through the mixing of the complex scalar with the 
standard model Higgs particle and the other portal comes from the interaction of the $U(1)'$ gauge boson, $Z'$, 
merely with the leptons in the standard model, hence being a leptophilic $Z'$ portal. 
One of the distinctive characteristics of our two-portal model is that the DM-nucleon elastic scattering
begins at one loop level. 
Therefore there is a large region in the parameter space which evades direct detection.
Thus, indirect detection searches become very important tools to probe the viable parameter space of the present model. 

In addition to the constraints from the relic density as well as the direct and indirect bounds on the dark matter model,
we examine the model if it is consistent also with the new observed DAMPE bump in the electron and positron flux 
in the cosmic rays. 

The paper have the following parts. In the next section we elaborate the setup of our leptophilic dark matter scenario. In 
section \ref{rel} the dark matter relic density and the invisible Higgs decay are computed and compared with bounds from 
the WMAP/Planck and the LHC. Next we take into account the muon magnetic anomaly and shrink the 
viable space of parameters. Constraints from the LEP is discussed in section \ref{LEP}. In section \ref{dir} we constrain
more the model with limits from the direct detection 
experiments specially the recent XENON1t and LUX experiments. 
Discussions on the neutrino trident production and $\tau$ decay is given in section \ref{trident}.
We also find a viable space of parameter consistent with the excess observed by the DAMPE in section \ref{dampe}. 
The Fermi-LAT constraint is discussed in section \ref{Fermi-LAT}.
Finally we conclude in section \ref{con}.

\section{Model}\label{mod}
We explore a leptophilic two-portal dark matter scenario. That is, a fermionic candidate of dark matter connected to
the standard model particles through vector and Higgs portals. The vector in the dark sector interacts with all
the lepton flavors in the SM but with no interaction with the quarks. The Lagrangian of the model 
can be written in three parts, 
\begin{equation}\label{lagtot}
\mathcal{L}=\mathcal{L}_{\text{SM}}+\mathcal{L}_{\text{DM}}+\mathcal{L}_{\text{int}},
\end{equation}
where the dark matter Lagrangian consists of a Dirac fermion playing the role of the dark matter 
and a complex scalar field both charged under $U\left(1\right)'$,
\begin{equation}\label{lagdm}
\begin{split}
 \mathcal{L}_{\text{DM}}=-\frac{1}{4} F'_{\mu\nu}F'^{\mu\nu}+\bar{\psi}\left(i\gamma^{\mu}D'_{\mu}-m_{\psi}\right)\psi \\
+\left(D'_{\mu}\varphi\right)\left(D'^{\mu}\varphi\right)^*-m^{2}(\varphi \varphi^*)
-\frac{1}{4}\lambda_s (\varphi \varphi^*)^2 \,.
\end{split}
\end{equation}
where the $U(1)'$ field strength is denoted by $F'_{\mu\nu}$, the $\psi$ is the Dirac fermion and $\varphi$ stands for the 
complex scalar. The dark sector covariant derivative is defined as,
\begin{equation}
D'_{\mu}=\partial_{\mu}- i g' z Z'_\mu. 
\end{equation}
which acts on the fields in the dark sector as well as the leptons in the SM with $g'$ being the strength of 
its coupling and $z$ the charge of the field acting on. 

Here we study a leptophilic model in which the $U(1)'$ gauge boson, $Z'$, interacts only with
the leptons in the SM but also with a hypothetical right-handed neutrino for the reason that will be discussed latter on. 
It is therefore necessary to modify the covariant derivative in the SM to include a term for the new coupling,
\begin{equation}
  D^{\text{SM}}_\mu  \rightarrow   D'^{\text{SM}}_\mu =   D^{\text{SM}}_\mu -i  g'  z  Z'_\mu \,,
\end{equation}
where $g'$ is the $U(1)'$ coupling in the dark sector and $z$ is the dark charge of the leptons that the covariant 
derivative acts on.

The interaction Lagrangian then reads, 
\begin{equation}\label{lint}
\begin{split}
 \mathcal{L}_{\text{int}}=
&-\lambda' (\varphi \varphi^*) \left(HH^{\dagger}\right)\\
&+ g' z_{E_L} Z'_{\mu}\bar{E}_L\gamma^{\mu}E_L
+ g' z_{e_R} Z'_{\mu}\bar{e}_R\gamma^{\mu}e_R \\
&+ g' z_{\nu_R} Z'_{\mu}\bar{\nu}_R\gamma^{\mu}\nu_R \,, 
\end{split}
\end{equation}
where $E_L$ and $e_R$ are respectively the three families of left-handed lepton doublets and right-handed lepton
singlets including the right handed neutrinos. 
Notice that we have considered universal charges for all families of the leptons, i.e. we have 
taken the $z_{e_L}$ to be the lepton $U(1)'$ charge for each family of left-handed lepton 
doublet and $z_{e_R}$ to be the $U(1)'$ charge for $e_R,\mu_R$ and $\tau_R$. 

Having introduced a new $U(1)'$ coupled to the dark matter and the chiral fermions in the SM, one must be careful 
about the triangle anomalies. In order to remove such anomalies we choose the charges in eq. (\ref{lint}) 
to take the following two leptophilic choices, \\

{\textbf A)} {$z_\mu\neq 0, z_{\nu_\mu} \neq 0 $ }

\begin{equation}\label{anfree1}
\begin{split}
&z_{e_L}= 2a,~~~~~  z_{\mu_L}=- a,~~~~~ z_{\tau_L}=- a \\
&z_{\nu_{e_L}}= 2a,~~~~~  z_{\nu_{\mu_L}}=- a,~~~~~ z_{\nu_{\tau_L}}=- a \\
&z_{e_R}=- 2a,~~~~~  z_{\mu_R}=\ a,~~~~~ z_{\tau_R}= a \\
&z_{\nu_{e_R}}=- 2a,~~~~~  z_{\nu_{\mu_R}}= a,~~~~~ z_{\nu_{\tau_R}}= a
\end{split}
\end{equation}

{\textbf B)} {$z_\mu = 0, z_{\nu_\mu}=0$}
\begin{equation}\label{anfree2}
\begin{split}
&z_{e_L}= a,~~~~~z_{\mu_L}=0,~~~~~z_{\tau_L}=- a \\
&z_{\nu_{e_L}}= a,~~~~~z_{\nu_{\mu_L}}=0,~~~~~z_{\nu_{\tau_L}}=- a \\
&z_{e_R}=- a,~~~~~z_{\mu_R}=0,~~~~~z_{\tau_R}= a \\
&z_{\nu_{e_R}}=- a,~~~~~z_{\nu_{\mu_R}}=0,~~~~~ z_{\nu_{\tau_R}}= a
\end{split}
\end{equation}
where $a$ is a real number. In the choice {\textbf A} it is assumed that the charge of the lepton $\mu$ and that 
of its neutrino $\nu_{\mu}$ are non-zero while in the choice {\textbf B}
we set $z_\mu=z_{\nu_\mu}=0$. We will clarify latter on the reasoning for these choices. 
The existence of the right-handed neutrinos are crucial; without them the triangle anomalies can not be fixed. The 
charge of the dark matter Dirac fermion, $z_\psi$,  suffices to have opposite values for its left-handed 
and right-handed components, i.e. $z_{\psi_L} = -z_{\psi_R}$.
Fixing $a=1$ and substituting the anomaly-free charges in eq. (\ref{anfree1}) and eq. (\ref{anfree2}) into eq. (\ref{lint}) 
we obtain two interaction Lagrangians, \\

{\textbf A)}
\begin{equation}\label{lint1}
\begin{split}
\mathcal{L}_{\text{int}}= &-\lambda' (\varphi \varphi^*) \left(HH^{\dagger}\right)\\
& - 2 g' Z'_{\alpha}\bar{e}\gamma^{\alpha}\gamma^5e 
 - 2  g' Z'_{\alpha}\bar{\nu}_e\gamma^{\alpha}\gamma^5 \nu_e\\
&  + g' Z'_{\alpha}\bar{\mu}\gamma^{\alpha}\gamma^5 \mu 
 +   g' Z'_{\alpha}\bar{\nu}_\mu\gamma^{\alpha}\gamma^5 \nu_\mu\\
&  +  g' Z'_{\alpha}\bar{\tau}\gamma^{\alpha}\gamma^5 \tau 
 +   g' Z'_{\alpha}\bar{\nu}_\tau\gamma^{\alpha}\gamma^5 \nu_\tau\,, 
\end{split}
\end{equation}

{\textbf B)}
\begin{equation}\label{lint2}
\begin{split}
\mathcal{L}_{\text{int}}= &-\lambda' (\varphi \varphi^*) \left(HH^{\dagger}\right)\\
& - g' Z'_{\alpha}\bar{e}\gamma^{\alpha}\gamma^5e 
 -  g' Z'_{\alpha}\bar{\nu}_e\gamma^{\alpha}\gamma^5 \nu_e\\
&  +  g' Z'_{\alpha}\bar{\tau} \gamma^{\alpha} \gamma^5 \tau 
 +   g' Z'_{\alpha}\bar{\nu}_\tau\gamma^{\alpha}\gamma^5 \nu_\tau. 
\end{split}
\end{equation}

As seen in eqs. (\ref{lint1}) and (\ref{lint2}) the vector boson $Z'$ couples to the leptons axially. 
Note that in eqs. (\ref{lint1}) and (\ref{lint2}) the fields $e, \mu$ and $\tau$ are all Dirac fermions. 
Let us turn back to scalars in the SM and in the dark sector. The Higgs potential as usual is composed of a quadratic and a 
quartic term which guarantees a non-zero {\it vev} for the Higgs field fixed by the experiment to be $v_h=246$ GeV.  
After the electroweak symmetry breaking we denote the Higgs doublet as $H^\dagger=(0~ v_h+h)$ where $h$ 
is the fluctuation around the {\it vev} being a singlet real scalar.
The complex scalar field, $\varphi$ in eq. (\ref{lagdm}) has two degrees of freedom out of which only one takes non-zero 
expectation value. The component that takes zero expectation value goes for the longitudinal part of the $Z'$ 
dark gauge boson. Therefore, 
\begin{equation}
 \varphi \to v_s + s \,,
\end{equation}
with $s$ being a real scalar which mixes with the SM Higgs and $v_s$ the vacuum expectation value of the scalar $\varphi$. 
The Higgs portal interaction term in eq. (\ref{lint}) together with the scalar 
potential in eq. (\ref{lagdm}) and the Higgs potential at the {\it vev} of the scalars, 
leads to a non-diagonal mass matrix for the 
field space of $h$ and $s$. We diagonalize the mass matrix by rotating in the $h$ and $s$ space by the 
mixing angle $\theta$ (see \cite{Ghorbani:2015baa} for more details).  
After diagonalizing the mass matrix we end up with the physical masses that we denote by $m_h$, $m_{s}$.  
We are keeping
the same notations for the scalar fields $h$ and $s$ after the mixing. The couplings of the model are $\lambda'$
in eq. (\ref{lint}), the Higgs quartic coupling $\lambda_h$ in the Higgs potential, and the scalar quartic coupling $\lambda_s$ 
in eq. (\ref{lagdm}). 
These couplings are all expressible in terms of the physical masses, $m_h, m_{s}$ and the mixing angle $\theta$, 
\begin{equation} \label{couplings}
\begin{split}
\lambda_{h}  &= \frac{m^{2}_s \sin^2 \theta +m^{2}_{h} \cos^2 \theta }{2v^{2}_h}\,,\\
\lambda_s  &= \frac{m^{2}_s \cos^2 \theta +m^{2}_{h} \sin^2 \theta }{v_s^2/2}
      - \frac{v^2_h}{v_s^2} \lambda'  \,,\\
\lambda' &= \frac{m^{2}_{h}-m^{2}_s}{2\sqrt{2}v_h v_s} \sin 2\theta\,.
\end{split}
\end{equation}
The vacuum stability conditions on the potential already give rise to the following constraints on the couplings, 
\begin{equation}
\begin{split}
 \lambda_{h} &> 0\,,\\
 \lambda_s {v_s}^2 &> \lambda' {v_h^2}\,, \\
 {v_s}^2 (\lambda_{h}\lambda_s -2 \lambda'^2) &> v_h^2 \lambda' \lambda_{h}\,.
\end{split}
\end{equation}
The free parameters of the model can then be assigned as $m_\psi, m_{s}$, $\theta$, $v_s$ and $g'$.

\begin{figure}
  \includegraphics[angle=-90,scale=.32]{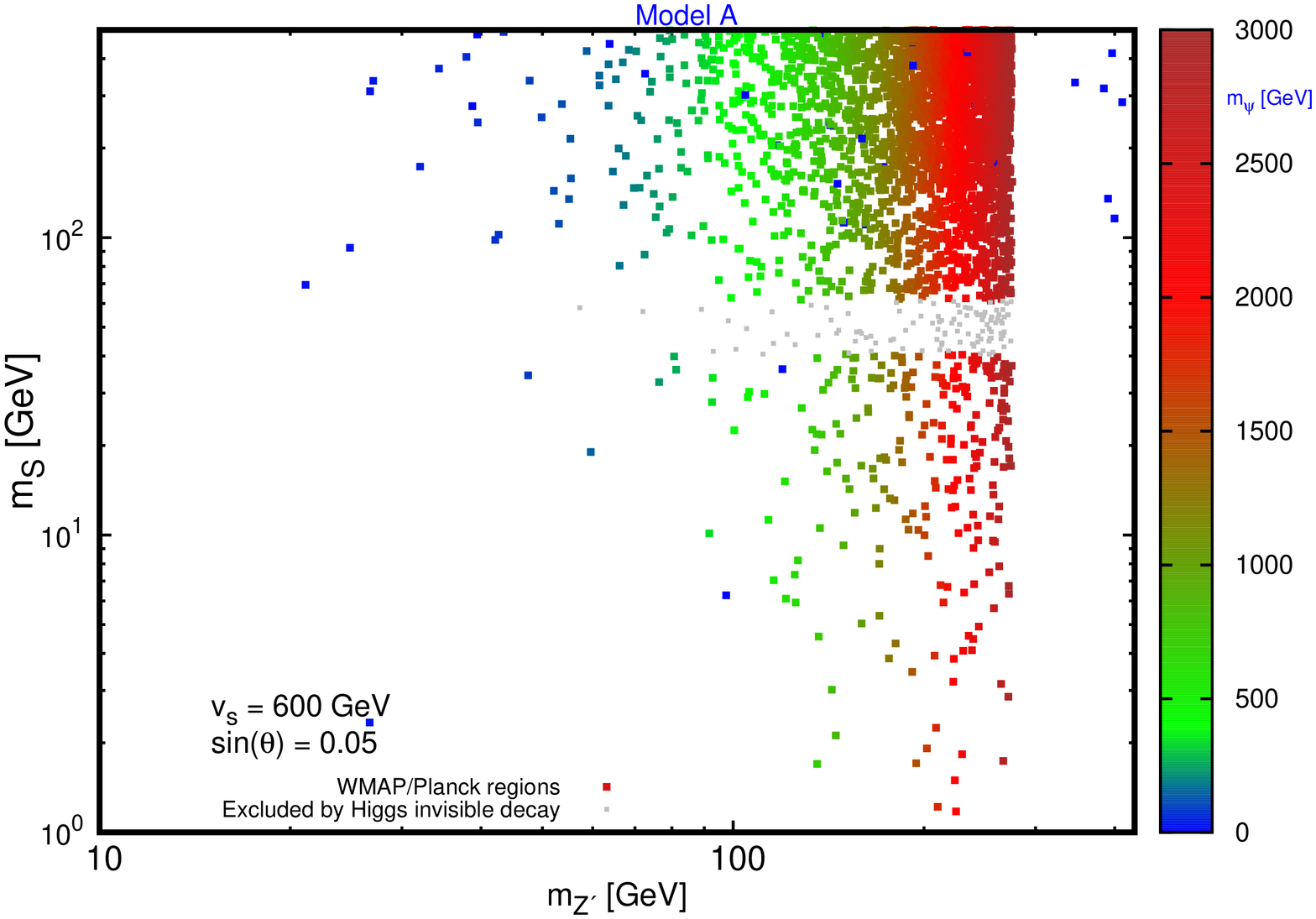}
   \includegraphics[angle=-90,scale=.32]{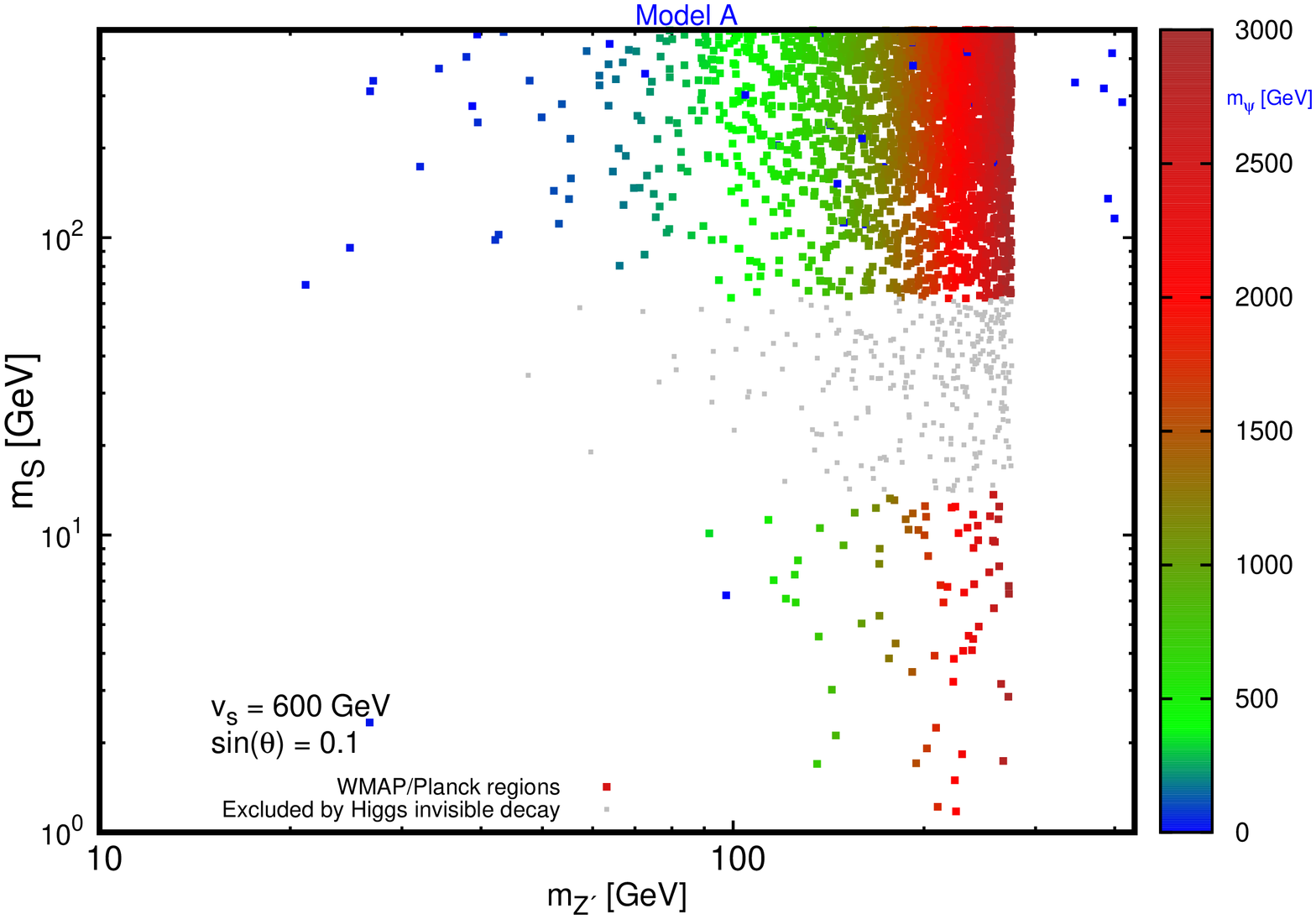}
   \caption{The plots show the mass of the $Z'$ against the mediator mass in model {\textbf A} for two mixing angles 
   {\it left)} $\sin \theta=0.05$
   and {\it right)} $\sin \theta=0.1$. The gray points are excluded by the invisible Higgs decay bound. The scan
   is done over the parameters with $10^{-3} < g^\prime < 1$, 1 GeV $< m_s <$ 500 GeV and
   10 GeV $< m_{\psi} <$ 3 TeV. In the scans $v_s = 600$ GeV.}
   \label{relA}
  \end{figure}

\section{Relic Density and Invisible Higgs Decay}\label{rel}
The fermionic DM candidate in the present model is a weakly 
interacting massive particle (WIMP). The basic vertices to build 
the diagrams relevant for the annihilation processes are as follows. 
The DM has a vector-type interaction with $Z'$, i.e., via 
the vertex $Z'_\mu \bar \psi \gamma^\mu \psi $, and the new gauge boson 
has axial-vector interactions with the SM leptons, i.e., 
via the vertex $Z'_\mu \bar l \gamma^5 \gamma^\mu l $.
Moreover, there are two types of vertices for the $Z'$ 
coupled to the SM Higgs and the new scalar, i.e., $Z'^\mu Z'_{\mu} h$ and $Z'^\mu Z'_{\mu} s$.
It is therefore possible to have DM annihilation in s-channel 
via $Z'$ exchange, $\bar \psi \psi \to \bar e e , \bar \mu \mu, \bar \tau \tau, \bar \nu_l \nu_l, Z' h, Z' s $ (for model \textbf{B},
$\bar \psi \psi \to  \bar\mu \mu$, $\bar \nu_\mu \nu_\mu$ are absent),
as well as in t- and u-channel with a DM exchange, $\bar \psi \psi \to Z' Z' $.
At temperature higher than the DM mass, the SM particles and the DM candidate are 
in thermal equilibrium based on the freeze-out paradigm. When the Universe expands 
the temperature cools down and as a consequence, the DM annihilation 
rate slows down. There is a temperature we call $T_{f}$ much below the DM mass 
where the DM annihilation rate drops right below 
the Hubble expansion rate of the Universe. At this time, 
the DM annihilation and production processes are suppressed and the number density 
of the dark matter remains constant afterwards. The dynamics 
behind the DM number density evolution is governed by the Boltzmann equation.
To obtain the current DM relic density, we make use of the package 
{\tt micrOMEGAs} \cite{Belanger:2001fz,Barducci:2016pcb}
which solves the equation numerically.
The DM annihilation cross sections are computed in {\tt CalcHEP} \cite{Belyaev:2012qa}.

\begin{figure}
   \includegraphics[angle=-90,scale=.32]{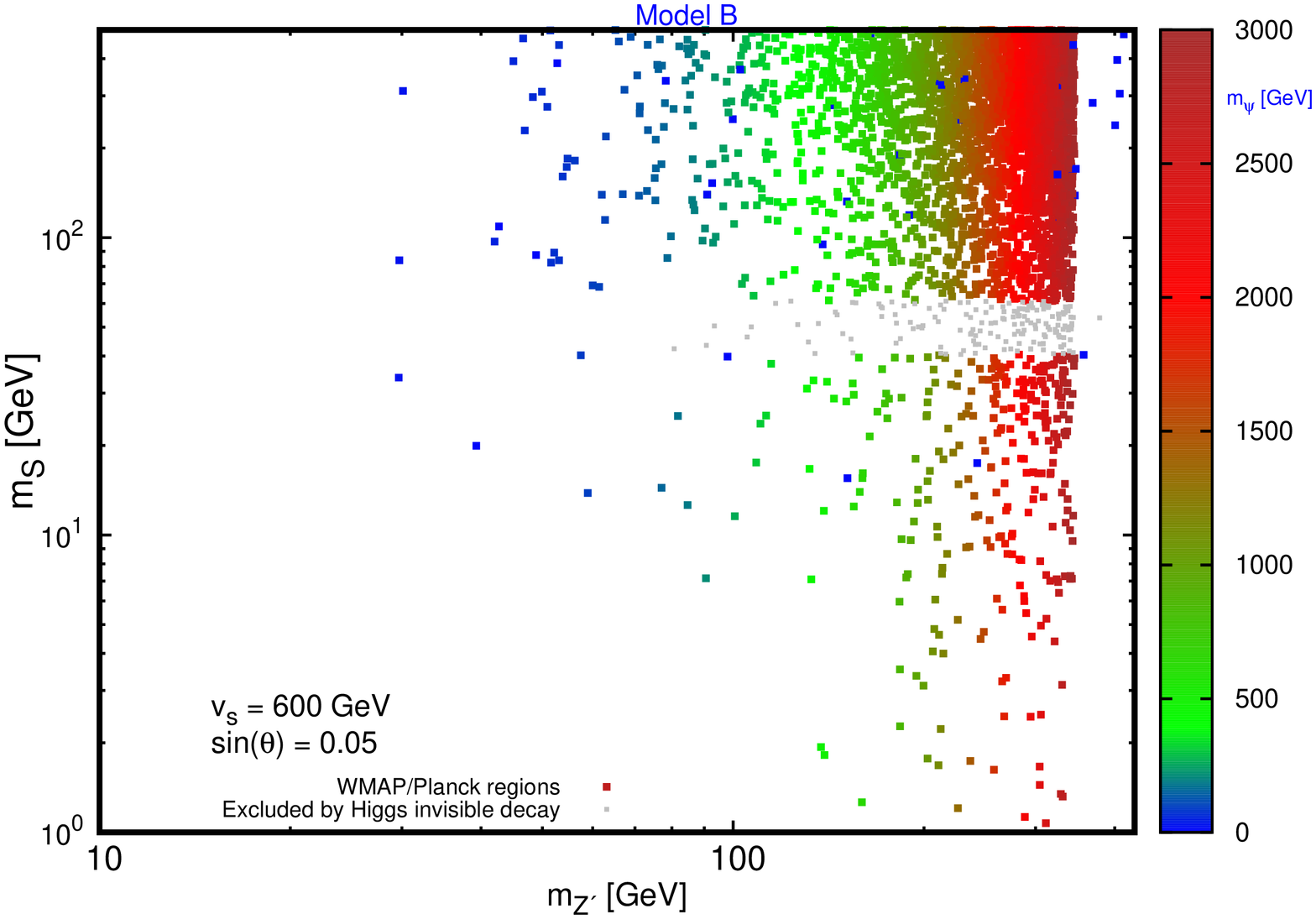}
  \includegraphics[angle=-90,scale=.32]{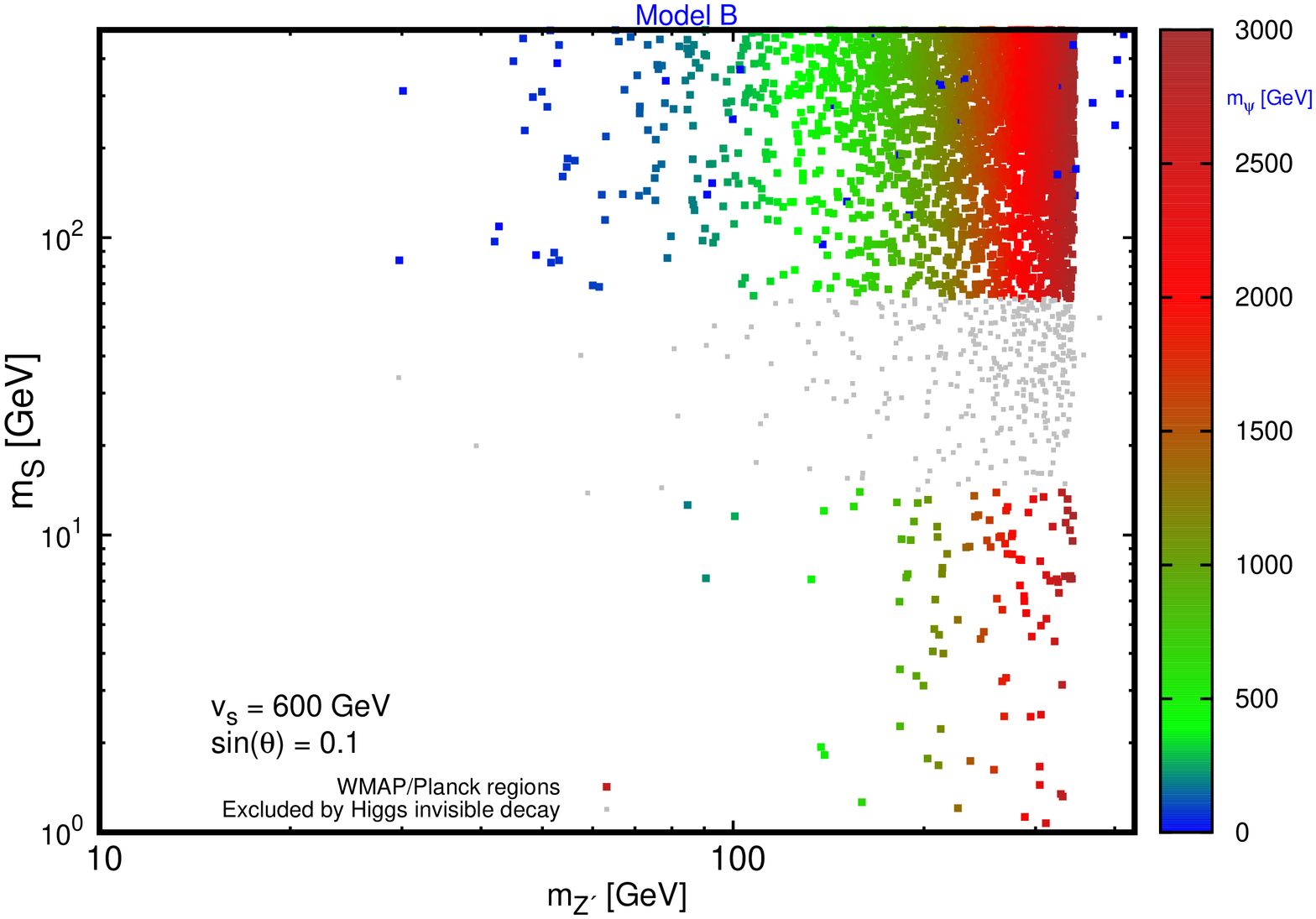}
     \caption{The plots show the mass of the $Z'$ against the mediator mass in model {\textbf B} for two mixing angles 
   {\it left)} $\sin \theta=0.05$
   and {\it right)} $\sin \theta=0.1$. The gray points are excluded by the invisible Higgs decay bound. The scan
   is done over the parameters with $10^{-3} < g^\prime < 1$, 1 GeV $< m_s <$ 500 GeV and
   10 GeV $< m_{\psi} <$ 3 TeV. In the scans $v_s = 600$ GeV.}
   \label{relB}
 \end{figure}

To constrain the model parameters we apply the observed DM relic density 
$0.1172 < \Omega_{\text{DM}} h^2 < 0.1226$ \cite{Hinshaw:2012aka,Ade:2013zuv}.
In addition, to find the viable regions in the parameter space we impose limits on the 
Higgs invisible decay rate. In the current model, the SM Higgs can decay into $Z' Z'$ and $ss$  
if $m_{Z'} < m_{h}/2$ and $m_{s} < m_{h}/2$ respectively. The decay rate for $h\to Z' Z'$ is,
\begin{equation}
 \label{dark-decay1}
\Gamma^{\text{inv}}(h \to Z^\prime Z^\prime ) = \frac{ {v_s}^2 {g^\prime}^4 \sin^2\theta  }{16\pi m_{h}} 
(1-4 m^{2}_{Z^\prime}/m_{h}^2)^{1/2} \,,
\end{equation}
and for the decay $h \to s s$ it is, 
\begin{equation}
\Gamma^{\text{inv}}(h \to s s) = \frac{w^2}{128\pi m_{h}} (1-4 m^{2}_{s}/m_{h}^2)^{1/2},
\end{equation}
where $w$ is a function of the mixing angle and the couplings as,
\begin{equation}
\begin{split}
 w & = 6\sqrt{2} \lambda' v_s \sin^3 \theta 
     +12 \lambda_{h} v_h  \cos \theta \sin^2 \theta \\
    &-6 \lambda' v_h  \cos \theta \sin^2 \theta  
    +2 \lambda' v_h  \cos \theta  \\
  & + 3 \sqrt{2} \lambda_s v_s \cos^2 \theta \sin \theta 
   - 4 \sqrt{2}  \lambda' v_s  \sin \theta. 
\end{split}
\end{equation}
The Higgs total decay rate will be modified as,
\begin{equation}
 \Gamma^{\text{tot}}_{\text{Higgs}} = \cos^2 \theta \, \Gamma^{\text{SM}}_{\text{Higgs}} 
 +\Gamma^{\text{inv}}(h \to Z^\prime Z^\prime )+  \Gamma^{\text{inv}}(h \to s s) \,.
\end{equation}
The experimental total decay width of the Higgs obtained in the SM turns out to be $\Gamma^{\text{SM}}_{\text{Higgs}} \sim 4$ MeV.
We apply the upper limit on the Higgs invisible branching ratio, 
$\Gamma^{\text{inv}}_{\text{Higgs}} \lesssim 0.24$ \cite{Khachatryan:2016whc}.
Now for both choices {\textbf A} and {\textbf B} in eqs. (\ref{anfree1}) and (\ref{anfree2})
we take $v_{s} = 600$ GeV and generate random\footnote{Throughout the paper scan is carried out
  using uniformly distributed pseudo-random numbers with linear prior.}
  points in the parameter space with 
$0.001 < g' < 1$, 1 GeV $< m_s < 500$ GeV and 1 GeV $< m_\psi < 3000$ GeV.
In Figs.~\ref{relA} and \ref{relB} we have illustrated the regions in the parameter space for models {\textbf A} and {\textbf B}
respectively,
which respect the expected relic density and the regions that are excluded by 
the experimental limit on the Higgs invisible decay. 
The results are compared for two different mixing angles $\sin \theta = 0.05$ and $\sin \theta = 0.1$.
It is evident from the plots that the excluded regions by the invisible Higgs decay
depends strongly on the mixing angle. For the larger mixing angle the scalar masses 
in the range $10$ GeV $\lesssim m_{s} \lesssim 60$ GeV are excluded, while for the smaller mixing angle
scalar masses in the range $40$ GeV $\lesssim m_{s} \lesssim 60$ GeV are excluded. 
In both cases a wide range of the DM mass are found viable. 
The mixing angle is fixed at $\sin \theta = 0.05$ in our analysis hereafter.

\section{Muon Anomalous Magnetic Moment}\label{muo}

One of the most precisely measured quantity in physics is the muon anomalous magnetic moment. The recent 
experiment at the Brookhaven National Laboratory (BNL) provides us with its value \cite{Bennett:2006fi},
\begin{equation}
a_{\mu} = \frac{g_{\mu}-2}{2} = (116592080\pm63)\times 10^{-11}  \,. 
\end{equation}
In the theoretical side, the computation of this quantity is a rather cumbersome task which 
involves contributions from many processes in QED, QCD and electroweak sectors. 
Although, the theoretical prediction 
of this quantity in the SM is affected by some uncertainties 
in the hadronic low energy cross section and hadronic vacuum polarization,  
it does not seem possible to explain the observed deviation of around $3.6 \sigma$ 
when compared with the recent NBL data: $\Delta a_{\mu} (\text{Exp-SM}) = (29.5\pm 8.1) \times 10^{-10}$.
This deviation may originate from some unknown physics beyond the SM (the new physics) 
or it could equally arise from some unknown sources in the current physics. 
When we consider models beyond the SM to explain the shortcomings of the SM, 
the contribution of the new physics to the muon anomaly should respect the confined bound 
on $a_{\mu}$. 
In the present work, only the model {\textbf A} introduces an axial coupling of $Z^{\prime}$ to 
the muon and therefore can potentially contribute a sizable amount to $a_{\mu}$ as 
\begin{equation}
\Delta a_{\mu} = \frac{g^{\prime} m_{\mu}^2}{8\pi^2 M_{Z\prime}^2} 
\int^{1}_{0} dx \frac{2x(1-x)(x-4)-4\alpha^2 x^3}{1-x+\alpha^2x^2} \,,
\end{equation}
where $\alpha = m_{\mu}/m_{Z'}$ \cite{Jegerlehner:2009ry}.
In the limit where, $\alpha  \ll 1$, we find 
\begin{equation}
 \Delta a_{\mu} \sim -\frac{5 g'^{2} m_{\mu}^2}{12\pi^2 m_{Z'}^2}  \,,
\end{equation}
which means that the $Z'$ coupling to the muon makes a negative contribution to the muon anomaly.
Given the $Z'$ mass, $m_{Z'} = g' v_s/\sqrt{2}$, $\Delta a_{\mu}$ will then depends only on 
the free parameter $v_s$ as 
\begin{equation}
\Delta a_{\mu} \sim -\frac{5m_{\mu}^2}{6\pi^2 v_s^2}  \,.
\end{equation}
Here we will see that by applying the measured value for $\Delta a_{\mu}$, 
the parameter $v_s$ is constrained strongly such that 
$497~\text{GeV} < v_s < 659~\text{GeV}$. Note that there is no bound on the model {\textbf B}
from the muon magnetic anomaly.

\begin{figure}
 \includegraphics[angle=-90,scale=.32]{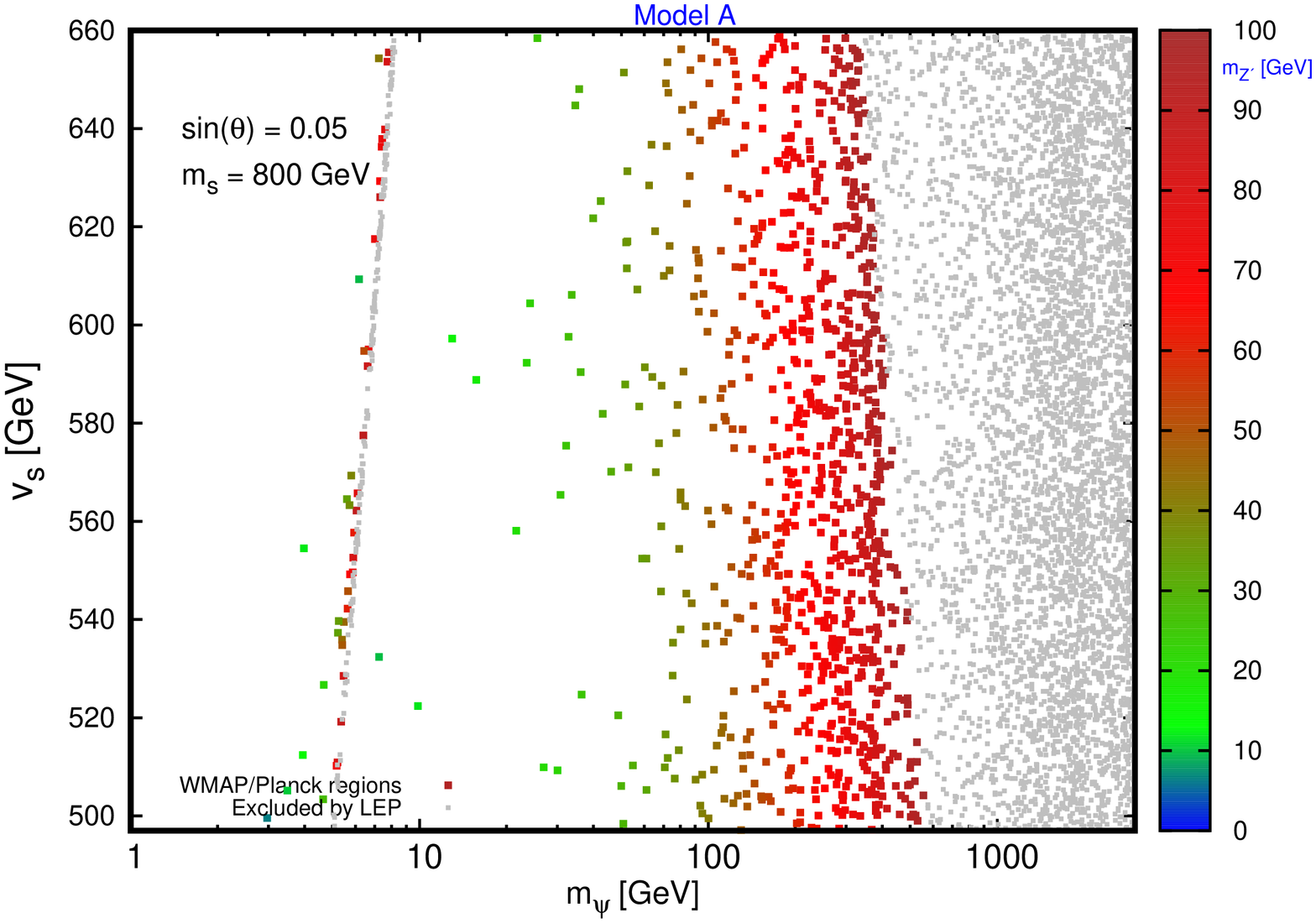}
 \includegraphics[angle=-90,scale=.32]{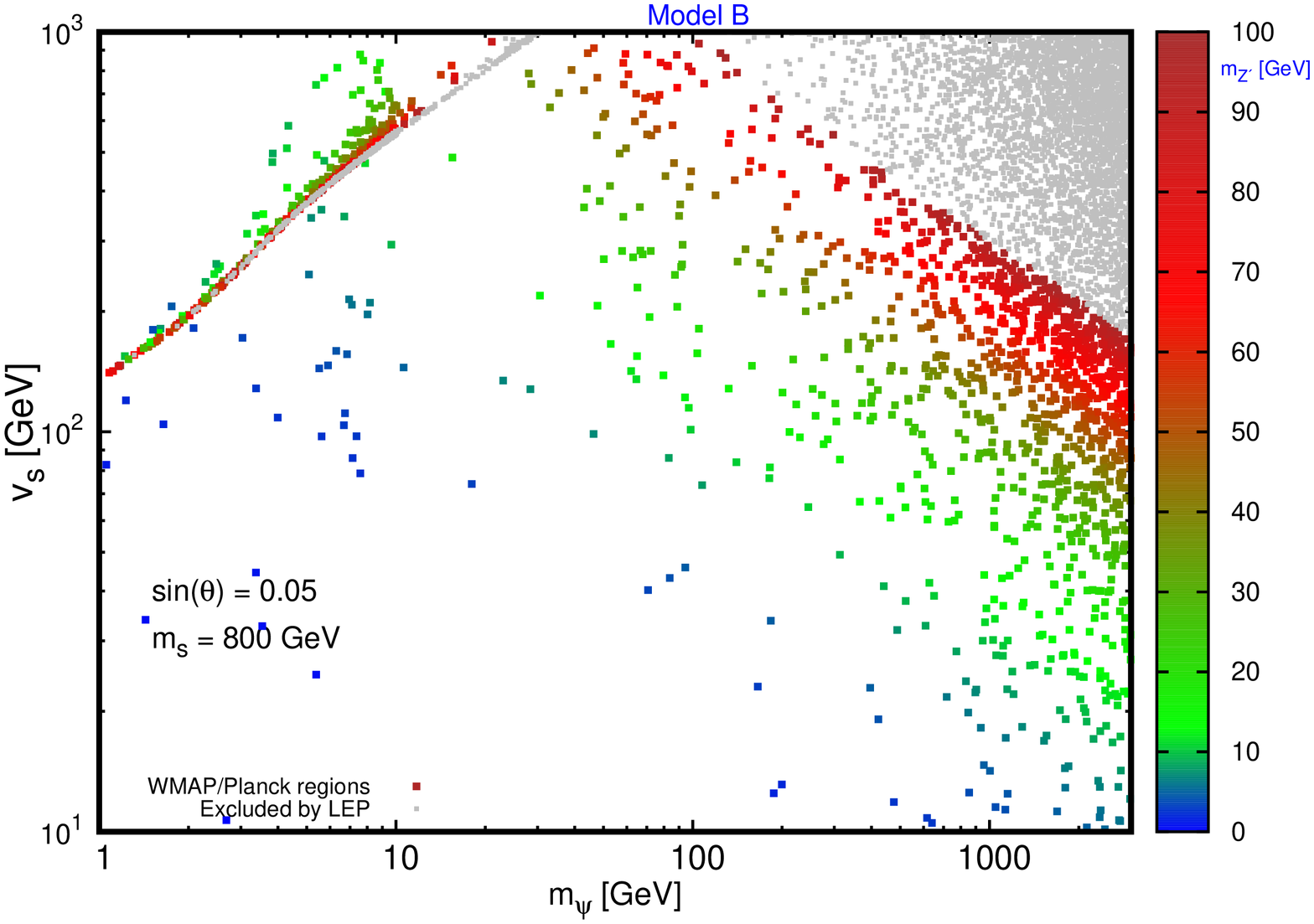}
 \caption{Comparing the viable region for the DM mass, $m_{\psi}$, and $m_{Z'}$ against the $v_s$ for {\it left)} model {\textbf A} and 
   {\it right)} model {\textbf B}. The gray region is excluded by the LEP. The muon anomaly is applied in the model {\textbf A}
   by taking $497$ GeV $<v_s<659$ GeV in the scan. The other parameters in the scan are 1 GeV $< m_{\psi} < 3$ TeV and $10^{-3} < g^\prime < 1$.
   The singlet scalar mass is fixed at $m_s = 800$ GeV.}
 \label{scanvev}
\end{figure}

\begin{figure}
\centering
 \includegraphics[angle=-90,scale=.35]{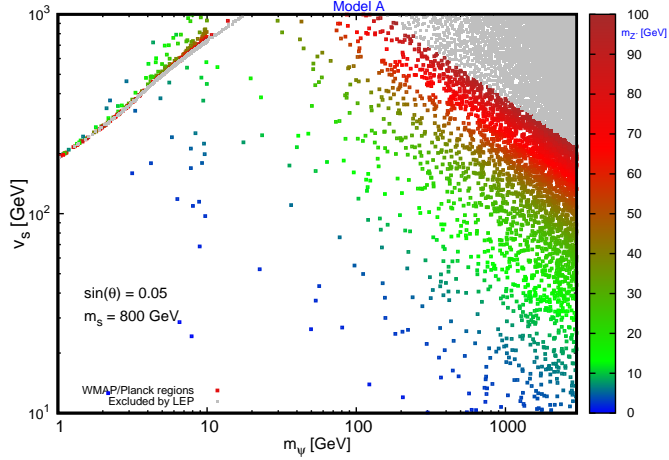}
 \caption{Same scan for model {\textbf A} as in Fig.~\ref{scanvev} except that the $(g-2)_\mu$ constraint on model A is removed.}
 \label{scanvev-modelA}
\end{figure}

\section{LEP constraint}\label{LEP}

Leptophilic dark matter models could be restricted by the results of the dismantled Large Electron-Positron Collider (LEP) 
in the $e^- e^+ \to e^- e^+$ scattering experiment (see e.g. \cite{LEP:2003aa,Freitas:2014jla}).  
In a model-independent four-fermion effective 
field theory framework investigation in \cite{Freitas:2014jla} the LEP 
puts constraint on  $g'$, the $Z'$ coupling to the electron in eqs. (\ref{lint1}) and (\ref{lint2}) as, 
\begin{equation}\label{leplim}
 \begin{split}
 & g'/m_{Z'}< 2.4 \times 10^{-4}\,\text{GeV}^{-1} ~ (m_{Z'}\gtrsim 200‍‍‍~\text{GeV})\\
 & g'/m_{Z'}< 6.9 \times 10^{-4}\,\text{GeV}^{-1} ~ (100~\text{GeV} \lesssim m_{Z'} \lesssim 200~\text{GeV})\,.
 \end{split}
\end{equation}
We note that the mono-photon constraint from LEP is sensitive to light DM mass \cite{Fox:2011fx}.
The benchmark for our DM mass is $m_{\text{DM}} \sim 1.5$ TeV. 
The LEP mono-photon constraint is not relevant since our DM mass is well 
above the maximum LEP center of mass energy.

For both models {\textbf{A}} and {\textbf{B}} in the current work we have imposed the LEP limits in eq. (\ref{leplim}). 
When considering also the relic density, the invisible Higgs decay and the muon anomaly bounds, the resulting viable space
is shown in the Fig. \ref{scanvev}. As seen in this figure, for the case \textbf{A} where the muon anomaly selects
out the $v_s$ to be only in the range $497~\text{GeV} < v_s < 659~\text{GeV}$, 
the DM mass is shrunk into $m_\psi\lesssim 550 $ GeV.
However for the model {\textbf{B}} where the muon anomalous magnetic moment is not restrictive the DM mass, $m_\psi$, can take 
values greater than $1.5$ TeV if $v_s\lesssim 200$ GeV.  For both cases the scalar mass and the mixing angle are fixed at
$m_s=800$ and $\sin\theta=0.05$, respectively. 
If we relax the $(g-2)_\mu$ constraint on model \textbf{A}, as shown in Fig.~\ref{scanvev-modelA} the 
viable parameter space of model \textbf{A} becomes similar to that of model \textbf{B}.

As will be discussed in section \ref{dampe} it is only the model {\textbf{B}} 
that can be tested against the recently observed DAMPE excess. In Fig. \ref{scanvev} the range of the $Z'$ mass 
has also been shown in color spectrum. It is evident from the figure that the large DM masses can be 
produced by either very light $Z'$ or heavier ones until $m_{Z'}=100$ GeV.

\section{Direct Detection}\label{dir}
We consider two types of scattering for the DM in our discussions 
about direct detection experiments; one is the nucleon-DM scattering 
and the other one is the DM scattering off the atomic electrons. Let us recall that
the DM candidate in our model has a vector interaction with $Z'$, while $Z'$ has an axial-vector
coupling to the SM leptons and no coupling to SM quarks.  

Assuming that non-relativistic DM with the mass $m_{\psi}$ scatters off the atomic electrons, 
the electron may be kicked out of the target atom. The elastic scattering cross section at tree-level then reads, 
\begin{equation}
\sigma_{\psi e} \sim \frac{g'^4 v^2_{\text{dm}} m_{e}^2}{2m_{Z'}^4}\,, 
\end{equation}
where the suppression factor $v_{\text{dm}}$ is the DM velocity in our galactic halo of order $\sim 10^{-3}$.
If we plug in the $Z'$ mass the cross section will depend only on $v_s$ as a free parameter, i.e.,  
$\sigma_{\psi e} \sim 2 v^2_{\text{dm}} m_{e}^2 / v^4_{s}$.
The XENON100 experiment results in null result for such a signal, however it puts an upper limit
on the elastic cross section as $\sigma_{\psi e} < 10^{-34} \text{cm}^2 (<100~\text{pb})$ \cite{Aprile:2015ade}.
This is a rather weak upper limit and as we will see cannot constrain the model parameters.

Now we turn into the nucleon-DM elastic scattering. In the present model 
this type of scattering can take place via loop induced Feynman diagrams because we deal with 
a leptophilic DM candidate. 

Since the SM Higgs and the scalar both interact with quarks (due to the mixing) and the $Z'$ boson,
one type of relevant Feynman diagram for the nucleon-DM elastic scattering 
is possible as depicted in Fig.~2 in \cite{Ghorbani:2015baa}. In the computation of the scattering amplitude,
we use the limit $t \ll m_{\psi}, m_{Z'}$ for the momentum transfer. 
This is reasonable because for a xenon nucleus for instance, 
we have $t \sim 2 \times 10^{-3}$ $\text{GeV}^2$ \cite{Ghorbani:2015baa}. 
The final result for the spin-independent (SI) elastic scattering in terms of 
the reduced mass of the nucleon-DM, $\mu_{\psi N}$, reads, 
\begin{equation}
\sigma^{\text{N}}_{\text{SI}} = 
\frac{4 \alpha_{N}^2 \mu_{\psi N}^2}{\pi} \,,
\end{equation}
where,
\begin{equation}
 \alpha_{N} = m_{N} \Big( \sum_{q = u,d,s} F^{N}_{Tq} \frac{\alpha_{q}}{m_{q}} 
+ \frac{2}{27} F^{N}_{Tg} \sum_{q = c,b,t}   \frac{\alpha_{q}}{m_{q}} \Big) 
\end{equation}
contains the low energy form factors 
$F^{N}_{Tq}$ and $F^{N}_{Tg}$ \cite{Belanger:2013oya}, 
and 
\begin{equation}
\begin{split}
\alpha_{q} & = \frac{g'^4 v_s m_{q}}{4\pi^2m_{\psi} v} \times \Large[ \frac{\cos^2 \theta}{m^2_s}-\frac{\sin^2 \theta}{m^2_h} \Large] \times
\\
& [-2 +\gamma \log \gamma  - \frac{\gamma^2-2\gamma-2}{\sqrt{\gamma^2-4\gamma}}
\log \frac{\sqrt{\gamma} + \sqrt{\gamma-4}}{\sqrt{\gamma} - \sqrt{\gamma-4}}]\,,
\end{split}
\end{equation}
with $\gamma = (m_{Z'}/m_{\psi})^2$.

Another type of loop induced Feynman diagram which may contribute to the
nucleon-DM scattering is the one with charged leptons running in the loop.
The lepton loop is connected in one side to the quark current by a photon 
or a $Z$ boson exchange and in the other side to the DM current by a $Z'$ exchange.
The insertion of the $l \gamma^5 \gamma^\mu \bar l Z'_{\mu}$ vertex in the lepton
loop turns the integral over the lepton momentum into the form, 
\begin{equation}
  \int \frac{dq^4}{(2\pi)^4} \text{Tr} 
  \Big[\gamma^5\gamma^\mu \frac{k_\nu \gamma^\nu +m_l}{k^2-m^2_l} \frac{q_\nu\gamma^\nu+m_l}{q^2-m^2_l} \Big] \,,
\end{equation}
which is zero due to the odd number of $\gamma^5$ in the trace.
Therefore, this process has no effect on the nucleon-DM elastic scattering. 

\begin{figure}
 \includegraphics[angle=-90,scale=.32]{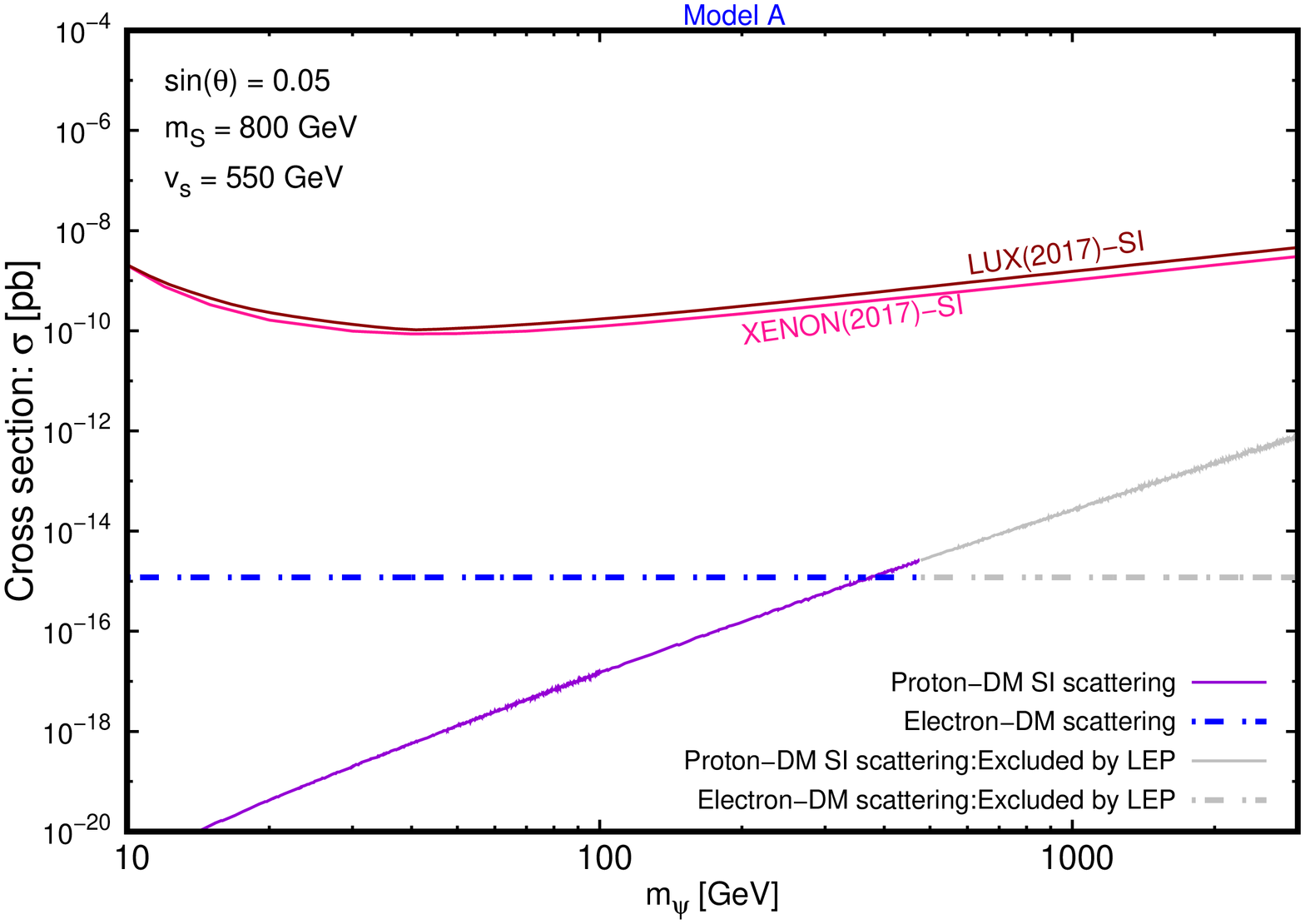} 
 \includegraphics[angle=-90,scale=.32]{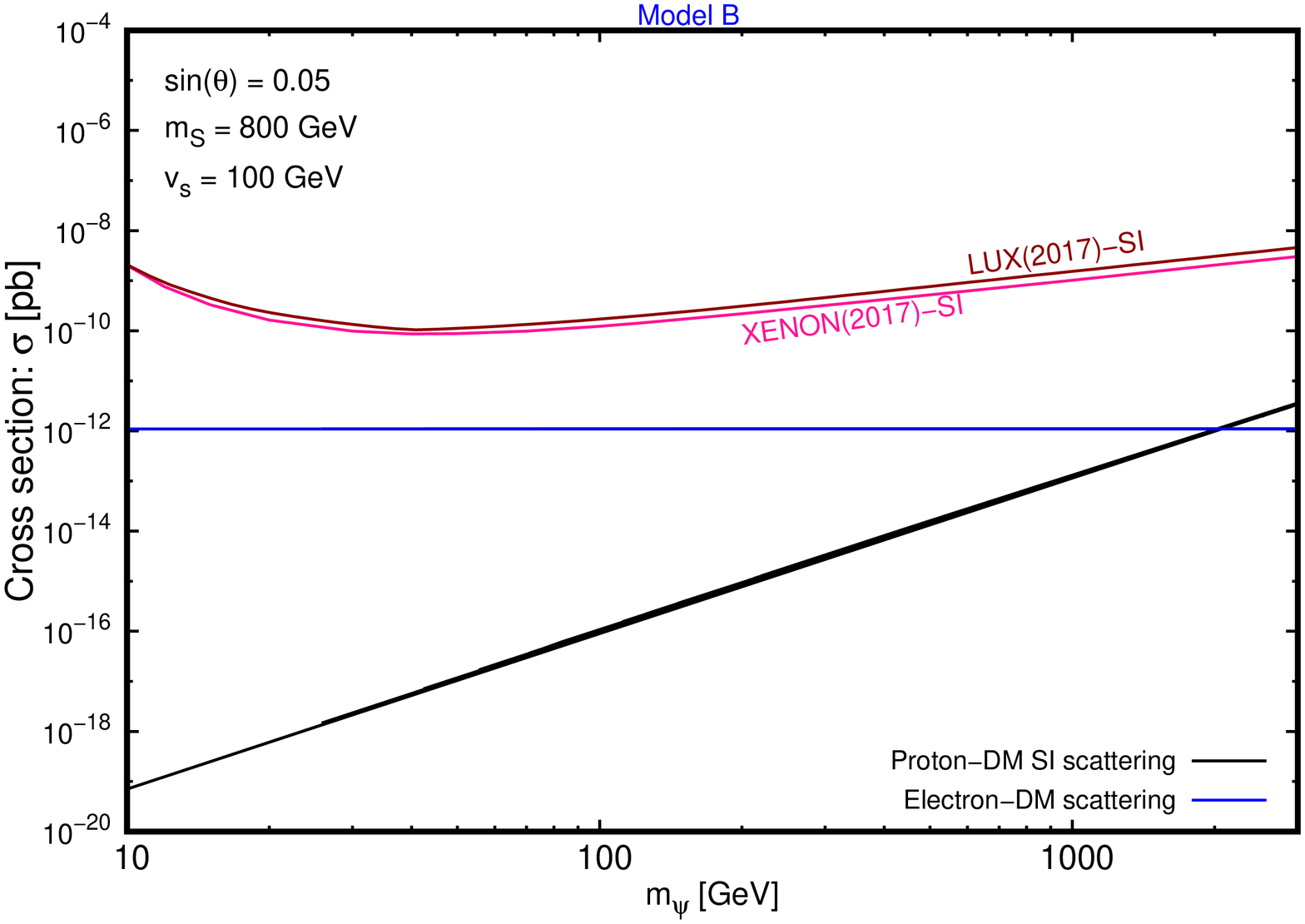}
  \caption{Cross sections for nucleon-DM scattering and electron-DM scattering are shown in terms 
 of the DM mass with $m_s = 800$ GeV and $\sin \theta = 0.05$ and results are compared for {\it left)} 
 model \textbf{A} and {\it right)} model \textbf{B}. In both figures the relic density is consistent with
 the observed value.
 Upper limits  on the SI cross section provided by the XENON1t and LUX are imposed. The gray region is excluded by the 
 LEP. The gray region is excluded by the LEP. The muon anomaly is applied in model \textbf{A}
 by taking $497$ GeV $<v_s<659$ GeV in the scan. Other parameters in the scan are 10 GeV $< m_{\psi} < 3$ TeV and
 $10^{-3} < g^\prime < 1$. The relic density constraint is applied here.}
 \label{DD}
\end{figure}

We scan over the parameter space while the mixing angle is fixed at $\sin \theta = 0.05$ and,
to satisfy the constraint from the muon anomalous magnetic moment we choose $v_{s} = 550$ GeV 
for the model {\textbf{A}}. It is chosen $v_{s} = 100$ GeV for the model {\textbf{B}}.
In both models we choose $m_s = 800$ GeV.
It is found out from the results in Fig.~\ref{DD} that for both models,  
DM masses up to 3 TeV respect the upper limits on the nucleon-DM cross section
imposed by the experiments XENON1t \cite{Aprile:2017iyp} and LUX \cite{Akerib:2016vxi}.
However, DM masses larger than $\sim 500$ GeV are excluded by the LEP in the model \textbf{A}.
With the same set of fixed parameters we also compute the electron-DM cross section. 
Since we have fixed $v_s$ at our analysis and the cross section depends on the this 
free parameter only, the cross section shows the same behavior in the plots in Fig.~\ref{DD} 
and its magnitude in both models is pretty much suppressed and resides well below the upper limit imposed by 
the XENON100. 
As seen in Fig. \ref{DD}, the dark matter mass in model \textbf{B} 
can take a large range of values from a few GeV to a few TeV 
after taking into account all the constraints discussed so far.

\begin{figure}
\centering
 \includegraphics[angle=-90,scale=.35]{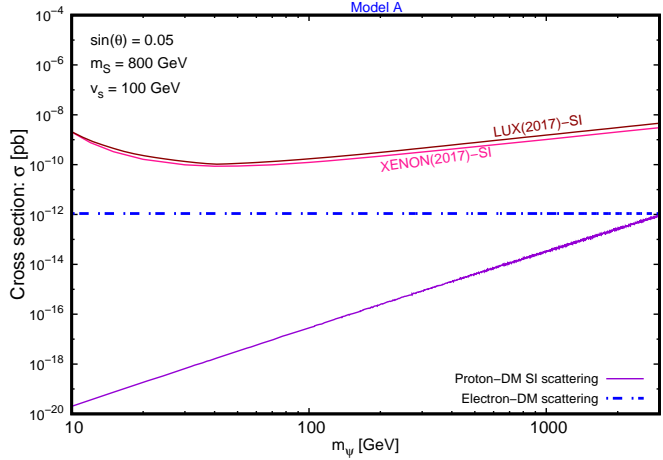} 
  \caption{Same scan for model \textbf{A} as that in Fig.~\ref{DD}, except that the $(g-2)_\mu$ constraint 
  on the model \textbf{A} is relaxed.}
 \label{nog-DD}
\end{figure}

We have also examined the case where the $(g-2)_\mu$ constraint is removed for model {\textbf{A}}. 
In Fig.~\ref{nog-DD} we show our results where the set of parameters in the scan are the same as 
those for model {\textbf{B}}. In this case, we find that the viable parameter space respecting the upper 
limits from direct detections is almost the same as that of model {\textbf{B}}.

\section{Neutrino Trident Production and $\tau$ Decay}\label{trident}
Generally, neutrino trident production can constrain models with $Z^\prime$ coupling
to both $\mu$ and neutrino \cite{Altmannshofer:2014pba}. 
It restricts the $Z^\prime-\mu$ coupling, $g^\prime$, 
to values given by  $g^\prime \lesssim \frac{m_{Z^\prime}}{1 \text{TeV}}$.
For model {\textbf{A}}, at our benchmark point with $m_{\text{DM}} \sim 1.5$ TeV, we have $m_{Z^\prime} \sim 32$ GeV 
while the relevant coupling is $g^\prime \sim 0.45$. 
Therefore the neutrino trident production excludes our benchmark point in model {\textbf{A}}.

Moreover, according to the results in \cite{Altmannshofer:2014cfa} for $\tau$ decay to muons, the region of parameter space 
at $m_{Z^\prime} \sim 32$ GeV is restricted to couplings in the range $0.15 \lesssim g^\prime \lesssim 0.25$.
Therefore in model {\textbf{A}}, our benchmark point with $m_{\text{DM}} \sim 1.5$ TeV, $m_{Z^\prime} \sim 32$ GeV and $g^\prime \sim 0.45$
is excluded by the $\tau$ decay to muons.

\section{DAMPE Excess}\label{dampe}

The high energy cosmic-ray electrons and positrons (CREs) flux is measured with high resolution and low background 
by the DAMPE (DArk Matter Particle Explorer) in the range $25$ GeV-$4.6$ TeV. 
The electrons and positrons propagate through the interstellar space and the evolution
of their energy distribution, $f_e$, is governed by the equation
\begin{equation}
 \partial_{t} f_e - \partial_{E} (b(E) f_e) -D(E) \nabla^2 f_e = Q_e({\textbf x},E)  \,.
\end{equation}
In the above equation the energy loss coefficient is $b(E) = -dE/dt$ which is parametrized 
in terms of the energy as $b(E) = b_{0} (E/\text{GeV})^2$ with $b_{0} = 10^{-16}\, \text{GeV}\,s^{-1}$. 
The diffusion factor, $D(E)$, depends on the energy and the disk thickness, $2L$, in the $z$ direction of the 
diffusion zone. It is parametrized as $D(E) = D_{0} (E/\text{GeV})^{\delta}$ 
with $D_{0} = 11 \, \text{pc}^2 \,\text{kyr}^{-1}$ and $\delta = 0.7$. 
The last ingredient in the diffusion equation is the source function, $Q_{e}$, 
for electrons and positrons in the case of DM annihilation. For a Dirac DM  candidate
the source function is given by 
\begin{equation}
 Q_{e}({ \textbf x},E) = \frac{\rho({\textbf x})^2}{4 m^2_{\text{DM}}}  \braket{\sigma v} \frac{dN}{dE} \,,  
\end{equation}
where $\rho({\textbf x})$ is the DM mass density, $\braket{\sigma v}$ is 
the velocity-averaged annihilation cross section of the DM and the energy spectrum of $e^{\pm}$ 
per annihilation is denoted by $dN/dE$ (see \cite{Cirelli:2008id} for more details). 

In case that the energy distribution, $f_{e}$, is time independent, the general solution for 
the energy distribution is given by the integral,
\begin{equation}
 f_{e} ({\textbf x}, E) = \int_{E}^{m_{\text{DM}}} dE_{s} \int d^3 { \textbf x}_{s} G({\textbf x},E; {\textbf x}_{s},E_{s}) 
   Q({\textbf x}_{s},E_{s})  \,,
\end{equation}
where the space integration is performed over the region of the DM halo and $E_s$ is the energy at the source.
The Green function of the diffusion equation 
is denoted by $G({\textbf x},E; {\textbf x}_{s},E_{s})$ and understood as the probability to catch 
an electron or a positron at earth with energy $E$ which is produced at point ${\textbf x}_{s}$ 
and energy $E_s$ in the DM halo. Finally, the electron and positron flux per unit energy is
obtained as $\Phi_e (E)= v f_{e}(E)/(4\pi)$, where $v$ is the electron or positron velocity.

In this work, to explain the enticing peak in the electron plus positron flux 
observed by the DAMPE, we assume that there is 
a DM subhalo nearby with a distance $d_{s} = 0.17$ kpc and subhalo radius $r_{s} = 0.1$ kpc. 
For the DM mass density in the subhalo we apply the NFW density profile \cite{Navarro:1996gj} 
\begin{equation}
 \rho (r) = \rho_s \frac{(r/r_s)^{-\gamma}}{(1+r/r_s)^{3-\gamma}}.
\end{equation}
In our numerical computation for the flux the code {\tt micrOMEGAs} is applied. 
In order to explain the flux at the peak position of about $1.4$ TeV, 
we assume the DM annihilation with the mass $\sim 1.5$ TeV in the subhalo. 
We then pick a point in the viable parameter space $m_{\text{DM}} = 1.5$ TeV 
consistent with the observed relic density and all other constraints. 
When the scalar mass is fixed at $m_s = 800$ GeV and $v_s = 100$ GeV, 
the for this benchmark point $g^\prime \sim 0.57$ and $m_{Z^\prime} \sim 40$ GeV.

\begin{figure}
\centering
 \includegraphics[angle=-90,scale=.40]{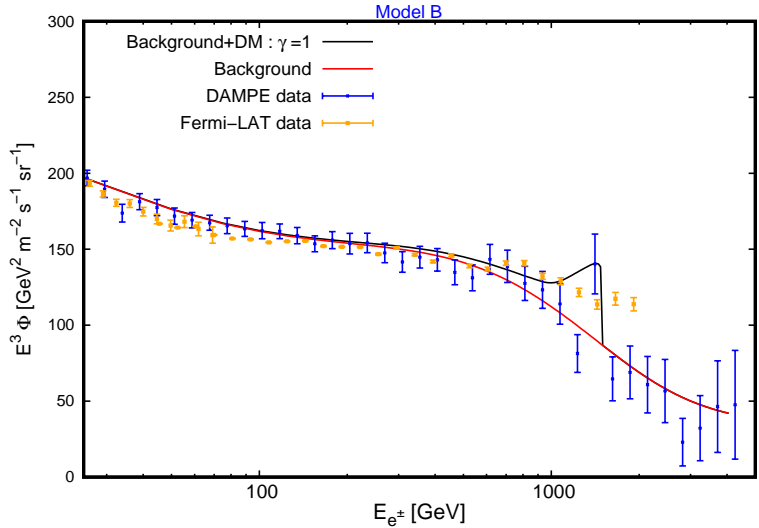}
  \caption{The electron and positron flux produced in a nearby DM subhalo is shown. A dark matter
  candidate with the mass about $1.5$ TeV plus the background explains the peak observed 
  by the DAMPE. The result is compared with the Fermi-LAT data for electron-positron flux.
  The relic density constraint is applied here, such that with the choices $m_s = 800$ GeV and $v_s = 100$ GeV
  it is obtained for the gauge coupling $g^\prime \sim 0.57$ and for the gauge boson mass $m_{Z^\prime} \sim 40$ GeV.}
 \label{flux}
\end{figure}

With the choice of the parameters as $\rho_s = 110~\text{GeV/cm}^3$ and $\gamma = 1$, we are 
able to explain the observed flux at $1.4$ TeV, as depicted in Fig.~\ref{flux}.
The cosmic ray (CR) background is computed in this work by following the formulas in \cite{Huang:2016pxg} 
for the primary electrons from the CR sources and the secondary electrons and positrons 
as a result of the primary electrons interaction with the interstellar medium.  
The relevant parameters in these formulas are obtained by 
the best fit using the electron plus positron flux measurement by the DAMPE \cite{Liu:2017rgs}.  
We also computed the thermally averaged dark matter annihilation cross section times the 
velocity at $m_\psi=1.5$ TeV. The result $\braket{\sigma v} \sim 2.2\times 10^{-26}$ cm$^3$/s
is compatible with the dark matter annihilation cross section predicted by the DAMPE. 
In Fig. \ref{flux} we included the Fermi-LAT electron-positron 
flux \cite{Abdollahi:2017nat} for comparison with the DAMPE data.

In principle the uncertainty on the parameters of the cosmic ray propagation may change our results.
For the benchmark point with DM mass $\sim 1.5$ TeV 
we checked this issue and realized that the deviation in our result is negligibly small.
This is also in agreement with the conclusions discussed in \cite{Belanger:2010gh}.

\section{Constraints from Fermi-LAT}\label{Fermi-LAT}
The Large Area Telescope (LAT) onboard the Fermi Gamma-ray Space Telescope, was the first to 
announce an excess in the gamma ray flux at a fairly low energy $\sim 2-3$ GeV. 
The DM annihilation at the Galactic Center was considered as a mechanism to explain the excess. 
However the latest finding by the Fermi Collaboration, indicates that the excess comes not only 
from the center of the galaxy but also from regions along the Galactic plane, where a DM signal
is not expected \cite{TheFermi-LAT:2017vmf}.
Given the assumption that dwarf spheroidal satellite galaxies (dSphs) accommodate 
a great deal of DM, the Fermi-LAT Collaboration could find the most strong limits on the 
cross section of DM annihilation into $\tau$ leptons (and $b$ quarks) by combined analysis of 15 dSphs in 
the Milky Way \cite{Ackermann:2015zua}.

In the present model, the DM annihilation into $\tau^+ \tau^-$ is the relevant channel. 
In this channel, the Fermi-LAT upper bound on the annihilation cross section is most sensitive 
to DM masses below 100 GeV. Therefore, a DM candidate of mass $\sim 1.5$ TeV can 
evade such upper limits. 

One important question is whether the Fermi-LAT have had the possibility
to detect a nearby $\gamma$-ray point source mimicking the DM subhalo we considered in
the present work to explain the DAMPE excess.
In the analysis reported in \cite{Yuan:2017ysv} the expected $\gamma$-ray fluxes are calculated
for a nearby clump with enhanced DM local density as the ones used to explain the DAMPE excess.
In the DM annihilation processes, the $\gamma$-ray may come along with electron-positron ($e^+ e^-$ channel),
or from internal bremsstrahlung processes and from decays of final state particles ($e\mu\tau$ channel).
In Ref.  \cite{Yuan:2017ysv} the Fermi-LAT isotropic background data are used to constrain the DM model.
It is found that only the $\gamma$-ray emission from the $e\mu\tau$ channels exceeds marginally
the Fermi-LAT upper limits, and $\gamma$-ray emission from other channels, i.e. $e^+e^-$ channel,
respect the Fermi-LAT constraints.

\section{Conclusion}\label{con}

The new observed electron-positron excess by the DArk Matter Particle Explorer (DAMPE) may open 
a window to new physics. The bump reported by the DAMPE in the electron-positron flux is interpreted from a $1.5$ TeV 
dark matter annihilation to electron-positron from a subhalo in about $0.1-0.3$ kpc away from the solar system. 
The dark matter annihilation cross section times the velocity must be of order $10^{-26}-10^{-24}$ cm$^3$/$s$.
To explain this excess
we introduce a model with a Dirac fermion as the dark matter candidate which has two portals to communicate with the 
SM, one way is through a complex scalar which mixes with the SM Higgs. And the other portal is through a $U(1)'$ gauge 
boson, $Z'$, interacting with the SM via only the leptons. The $U(1)'$ charges of the leptons and the dark matter Dirac 
fermion are chosen in a way to cancel the triangle anomalies. We have investigated two sets of charges once when the muon 
$U(1)'$ charge is vanishing and once it is non-zero. 
We then have computed the relic density and impose its value to be 
$\Omega_{\text{DM}} h^2 \sim 0.11$. By the bound from the invisible Higgs decay we restricted more the space of the parameters.
The LEP electron-positron collision results, restrict strongly the vacuum expectation value of the scalar and through which the 
masses of the $Z'$ and the DM. 
Considering all the bounds above we then have computed the DM-nucleus elastic scattering cross section and 
constrain the model by the recent direct detection 
experiments XENON1t/LUX. 
Constraints from Fermi-LAT observations, neutrino trident production and $\tau$ decay are also discussed.
The viable dark matter mass we obtain after imposing all the aforementioned limits contains a $1.5$ TeV 
dark matter mass which can produce an excess in the electron-positron flux matching the properties of the DAMPE excess.

 \appendix
 \section{Dark Matter Annihilation Cross Sections}
 \label{formula}
We provide the DM annihilation cross section formulas in this section for four different channels. 
First, the annihilation cross section for the annihilation process $\bar \psi \psi \to \bar f f$ 
with $f = l^+ l^-, \bar \nu_l \nu_l$ is obtained as
\begin{equation}\label{ff}
\sigma v_{\text{rel}} (\bar \psi \psi \to \bar f f) = 
\frac{{g^\prime}^4 \sqrt{1-4m_{f}^2/s}}{6 \pi s} 
\frac{(s^2-8m_{f}^2 m_{\psi}^2 + 2 s m_{\psi}^2 -\frac{4}{9}s m_{f}^2)}{(s-m^{2}_{Z'})^2+m^{2}_{Z'}\Gamma^{2}_{Z'}}
 \,.
\end{equation}

The other annihilation process is $\bar \psi \psi \to h Z'$, which is mediated 
by a $Z^\prime$ gauge boson via s-channel. We find the following result for 
the annihilation cross section as,  
\begin{equation}\label{z'h}
 \sigma v_{\text{rel}} (\bar \psi \psi \to h Z') = 
\frac{{g^\prime}^6 v_s^2 \sin^2 \theta}{16 \pi s} \frac{(s+2 m_{\psi}^2) \sqrt{[1-(m_{h}^2+m_{Z'}^2)/s]^2-4m_{h}^2m_{Z'}^2/s^2}}
{(s-m^{2}_{Z'})^2+m^{2}_{Z'}\Gamma^{2}_{Z'}} \,.
\end{equation}

Similarly, we get the DM annihilation cross section for the process $\bar \psi \psi \to s Z'$, 
\begin{equation}\label{z's}
 \sigma v_{\text{rel}} (\bar \psi \psi \to s Z') = 
\frac{{g^\prime}^6 v_s^2 \cos^2 \theta}{16 \pi s} \frac{(s+2 m_{\psi}^2) \sqrt{[1-(m_{s}^2+m_{Z'}^2)/s]^2-4m_{s}^2m_{Z'}^2/s^2}}
{(s-m^{2}_{Z'})^2+m^{2}_{Z'}\Gamma^{2}_{Z'}} \,.
\end{equation}

Finally, we find the annihilation cross section for the process $\bar \psi \psi \to Z' Z'$ 
with a DM particle as the mediator via t- and u-channel,
\begin{equation}\label{Z'Z'}
\begin{split}
 \sigma v_{\text{rel}} (\bar \psi \psi \to Z' Z') =
\frac{{g^\prime}^4 \sqrt{1-4m_{Z'}^2/s}}{8 \pi^2 s}
\int d\Omega \Big[ \frac{s m_{Z'}^2 - m_{\psi}^2m_{Z'}^2 + \frac{1}{2} s m_{\psi}^2-2m_{\psi}^4}{(t-m_{\psi}^2)(u-m_{\psi}^2)}\\
- \frac{(m_{\psi}^2+m_{Z'}^2-t)^2+ts -s m_{\psi}^2+2 t m_{\psi}^2+4 m_{\psi}^2 m_{Z'}^2+2 m_{\psi}^4 }
{2(t-m_{\psi}^2)^2}\\
- \frac{(m_{\psi}^2+m_{Z'}^2-u)^2+us -s m_{\psi}^2+2 u m_{\psi}^2+4 m_{\psi}^2 m_{Z'}^2+2 m_{\psi}^4 }
{2(u-m_{\psi}^2)^2} \Big] \,,
\end{split}
\end{equation}

where in the formulas above, $s$, $t$ and $u$ are the relevant mandelstam variables.

\bibliography{ref.bib}
\bibliographystyle{apsrev4-1}

\end{document}